\documentclass[aps,pra,twocolumn,showpacs,superscriptaddress]{revtex4-1}
\usepackage{amsmath,amssymb}
\usepackage{graphicx}
\usepackage{natbib}
\usepackage{bm}

\begin{document}

\title{Spin correlations and doublon production rate for fermionic atoms in modulated optical lattices}

\author{Akiyuki Tokuno}
\affiliation{DPMC-MaNEP, University of Geneva, 24 Quai Ernest-Ansermet
CH-1211 Geneva, Switzerland.}
\author{Thierry Giamarchi}
\affiliation{DPMC-MaNEP, University of Geneva, 24 Quai Ernest-Ansermet
CH-1211 Geneva, Switzerland.}

\date{\today}

\begin{abstract}
 We compute the integrated doublon production rate in response to a
 lattice modulation for two-component fermions in an optical lattice. 
 We derive a general formula for the integrated intensity, valid in the
 presence of inhomogeneous potentials such as the trap, which gives the
 integrated intensity in terms of equal time correlation functions
 only. 
 Such a formula is thus well suited for direct numerical calculations.
 We show that, in the limit of large repulsion for commensurate
 fillings, or for temperature ranges for which the hopping is
 incoherent, the integrated doublon spectrum is directly related to the
 nearest-neighbor spin-spin correlation function.
 We compute its temperature dependence in this regime using finite
 temperature quantum Monte Carlo calculation. 

\end{abstract}

%============================================================
%============================================================
\maketitle

Ultra-cold atom gases confined in an optical lattice provide a
controlled realization of the Hubbard
model~\cite{Bloch.Dalibard.Zwerger/RevModPhys80.2008:_review,Esslinger/AnuuRevCondMattPhys1.2010:review} 
which plays a central role in the study of strongly correlated electron 
systems such as Mott transition, high-$T_c$ superconductivity, and
quantum magnetism. 
Despite intensive research the phase diagram of this model is still
largely debated.  
Cold atoms, with the control of the interactions using a Feshbach
resonance technique and the control of kinetic energy via lattice depth,
allow us to probe such physics in detail. 

In order to do so, the development of experimental probes to capture
many-body quantum states is also an important issue. In connection with
the Hubbard physics, measuring the antiferromagnetic (AFM) correlations
is of prime importance.
However this is not an easy task. 
The AFM order can potentially be extracted from time-of-flight
measurements via shot-noise
measurement~\cite{Altman/PRA.70.2004,Greiner/PRL.94.2005,Foelling/Nat.434.2005}.
However, the measurement is far from trivial and also depends crucially
on the direction of the magnetic order. 
Local
addressing~\cite{Sherson.etal/Nature467.2010,Bakr.etal/Science329.2010} 
is also a potential route but systems are for the moment quite small,
and in addition such a technique is complicated to extend to
three-dimensional systems. 
It is thus highly desirable to study other probes that can potentially
give direct access to the magnetic
correlations~\cite{Weitenberg.etal/PRL106.2011,Trotzky.etal/PRL105.2010,Chen.etal/PRL107.2011,Pedersen.etal/PRA84.2011,Gorelik.etal/PRA85.2012,Gorelik.etal/arXiv.2011,Gorelik.etal/PRL105.2010}.

One probe which has proved to be very efficient and relatively simple to
implement is the amplitude modulation of the optical
lattice~\cite{Stoeferle.etal/PRL.92.2004}. 
This probe, first used for bosonic systems with a measurement of the
absorbed energy, gives direct access to the kinetic energy correlation
functions~\cite{Iucci.etal/PRA73R.2006,Reischl.etal/PRA72.2005,Kollath.etal/PRL97.2006}.
For fermions the energy absorption rate (EAR) cannot be implemented
accurately enough and a variant of this probe measuring the doublon
production rate (DPR) has been proposed~\cite{Kollath.etal/RPA74R.2006}.
This DPR measurement in the linear response regime could be successfully
implemented~\cite{Joerdens.etal/nature455.2008}, and thus stimulates
further studies of theoretical calculation of DPR
spectra~\cite{Sensarma.etal/PRL103.2009,Huber.Ruegg/PRL.102.2009,Xu.et.al/PRA84R.2011,Massel.Leskinen.Toermae/PRL103.2009,Korolyuk.Massel.Toermae/PRL104.2010,Tokuno.Demler.Giamarchi/PRA85.2012}
and doublon
dynamics~\cite{Sensarma.etal/PRB82.2010,Sensarma.etal/PRL107.2011,Stohmaier.etal/PRL104.2010}. 

In addition, it was shown~\cite{Kollath.etal/RPA74R.2006}
by a direct comparison of the two quantities, 
that the integrated intensity of the modulation does directly give
access to the nearest-neighbor AFM correlations.
Intuitively, for fermionic atoms, spin configurations of neighboring
atoms are relevant to the spectra because of the Pauli exclusive
principle: 
while the hopping is allowed for neighboring atom spins pointing in
anti-parallel, it is blocked for ferromagnetically aligned spin
configuration. 
This point, which allows one to use the modulation as a simple probe of
magnetism, was explored further both theoretically and experimentally in
the linear response regime at high
temperatures~\cite{Greif.etal/PRL.106.2011}. 

In this paper, we further analyze the integrated intensity of the DPR. 
We consider potentially inhomogeneous systems, e.g. due
to the presence of the trap.  
We show that the frequency integrated DPR can be fully expressed, within
linear response, in terms of static correlations. 
Such correlations, contrarily to the original DPR response at fixed
frequency, are well within the reach, without any need for analytical
continuation, of powerful numerical methods such as quantum
Monte Carlo (QMC) simulation,thus potentially allowing a very precise
comparison of the integrated DPR and theoretical calculations. 
For the case when the filling is commensurate and the interactions are
large compared to the kinetic energy, or when the hopping is incoherent
due to the temperature, that the above formulas reduce, in agreement
with the initial study of Ref.~\cite{Kollath.etal/RPA74R.2006}, to a
measure of the nearest-neighbor AFM correlation functions.
We obtain the temperature dependence of this quantity via a QMC
simulation of the Heisenberg quantum spin model in a three-dimensional
cubic lattice.

We consider two-component fermionic atoms strongly confined in an
optical lattice.
In the case of deep optical lattices, the physics is well described by
the spin-$1/2$ fermionic Hubbard model,
$H_0=H_{\mathrm{K}}+H_{\mathrm{U}}+H_{\mathrm{p}}$, with
\begin{align}
 % & H_{0}=H_{\mathrm{K}} + H_{\mathrm{U}} + H_{\mathrm{p}},
 % \label{eq:Hubbard-model}
 % \\
 & H_{\mathrm{K}}
 =-J\sum_{\sigma=\uparrow,\downarrow}\sum_{\langle{i,j}\rangle}
     c^{\dagger}_{i\sigma}c_{j\sigma},
 \label{eq:kinetic-energy}
 \\
 & H_{\mathrm{U}}
 = U\sum_{j}n_{j\uparrow}n_{j\downarrow},
 \\
 & H_{\mathrm{p}}
 =\sum_{j}v_{j}\left(n_{j\uparrow}+n_{j\downarrow}\right)
 \label{eq:interaction-energy}
\end{align}
where $c_{j\sigma}$ and $n_{j\sigma}$ are, respectively, an annihilation
and number operator of a fermionic atom at a $j$th site.
The on-site potential $v_{j}$ (for example, corresponding to trap
potential) generally breaks translational symmetry.

The parameters $J$ and $U$ are given as a function of an
amplitude of a sinusoidal lattice potential, i.e.,
$V_\mathrm{op}(\bm{r})=V_0\left[\cos^2(\pi x/a)+\cos^2(\pi y/a)+\cos^2(\pi z/a)\right]$,
where the lattice constant is
$a$~\cite{Bloch.Dalibard.Zwerger/RevModPhys80.2008:_review}.
The dynamical amplitude modulation defined on continuum space
is represented in an effective lattice model as follows:
$J\rightarrow J[1 + \delta{J}\cos(\omega t)]$ and
$U\rightarrow U[1 + \delta{U}\cos(\omega t)]$ corresponding to
$V_0\rightarrow V_0+\delta{V}\cos(\omega t)$ for the lattice
depth~\footnote{The second-order variations of $J$ and $U$ by the
modulation $\delta{V}$ can be taken into account as well, and do not
affect the results on the EAR and DPR given in this paper.}.
Thus the perturbation Hamiltonian of the lattice modulation is
given as
$V(t)=(\delta{J})\cos(\omega t)H_{\mathrm{K}}+(\delta{U})\cos(\omega t)H_{\mathrm{U}}$,
but here we use the alternative representation to simplify the
problem~\cite{Reischl.etal/PRA72.2005} as
$V(t)=(\delta{U})\cos(\omega t)H_{0}+\cos(\omega t)S$ with
\begin{equation}
 S=(\delta{F})H_{\mathrm{K}}-(\delta{U})H_{\mathrm{p}},
 \label{eq:modulation-perturbation}
\end{equation}
where $\delta{F}=\delta{J}-\delta{U}$ is a dimensionless perturbation
parameter.
Note that in the homogeneous case ($v_j=v$) the perturbation
operator becomes the kinetic energy $H_{\mathrm{K}}$.

In the linear response regime, the DPR per site can be defined as an
increment of the doublon number time-averaged over a single period
$2\pi/\omega$,
\begin{equation}
 P_{\mathrm{D}}(\omega)
 =\frac{1}{\Omega}
  \frac{1}{2\pi/\omega}
  \int_{t}^{t+(2\pi/\omega)}\!\! dt'\ \frac{d}{dt'}N_{\mathrm{D}}(t'),
 \label{eq:DPR-definition}
\end{equation}
where $N_{\mathrm{D}}(t)$ is the number of created doublons, and
$\Omega$ is a total site number of the system.
Note that as seen below, the time dependency in the right-hand side
should cancel due to the single-period time average in the linear
response region. 
The doublon number can also be written by the Hubbard interaction
$N_{\mathrm{D}}(t)=\langle{H_{\mathrm{U}}}\rangle/U$,
which is averaged by the density matrix
$\rho(t)=e^{-H(t)/k_{B}T}/\mathrm{Tr}[e^{-H(t)/k_{B}T}]$ for
time-dependent Hamiltonian $H(t)=H_0+V(t)$.
Thus the doublon number is also expressed as
\begin{equation}
 N_{\mathrm{D}}(t)
 =\frac{
   E(t)
   -\langle{H_{\mathrm{K}}+H_\mathrm{p}}\rangle
   -\langle{V(t)}\rangle}
   {U},
  \label{eq:doublon-number}
\end{equation}
where
$E(t)=\langle{H(t)}\rangle$ is a system energy.
In the second-order perturbation theory with respect to the lattice
modulation $V(t)$, all terms in Eq.~(\ref{eq:doublon-number}) apart from
the first one are found to generate only oscillatory terms whose
contribution to the DPR spectrum disappears due to the time average in
Eq.~(\ref{eq:DPR-definition}).
Therefore one can rewrite Eq.~(\ref{eq:DPR-definition}) as follows:
\begin{equation}
 P_{\mathrm{D}}(\omega)
 =\frac{1}{\Omega}
  \frac{1}{2\pi/\omega}
  \int_{t}^{t+(2\pi/\omega)}\!\! dt'\
    \frac{1}{U}\frac{d}{dt'}E(t').
 \label{eq:EAR}
\end{equation}
Interestingly, the equality means that the DPR is equivalent to the
EAR in the second-order response regime, as was pointed out in Ref.~\cite{Kollath.etal/RPA74R.2006}.
The EAR~(\ref{eq:EAR}) as a second-order response can be formulated by
the linear response theory.
As a result, the DPR formula can be
given~\cite{Kollath.etal/RPA74R.2006} as
\begin{equation}
 P_{\mathrm{D}}(\omega)
 =-\frac{1}{\Omega}\frac{1}{2\hbar U}
   \mathrm{Im}
     \left[
       \omega \tilde{\chi}^{\mathrm{R}}_{S}(\omega)
     \right],
 \label{eq:DPR-formula}
\end{equation}
where
$\tilde{\chi}^{\mathrm{R}}_{S}(\omega)=-i\int_{0}^{\infty} dt\,
e^{i\omega t}\langle[S(t),S(0)]\rangle_0$
is the Fourier transform of the retarded correlation function of the
operator $S$ for the unperturbed Hamiltonian $H_0$.

We now consider the general formula of DPR spectra
(\ref{eq:DPR-formula}) integrated over the modulation frequency $\omega$.
Before implementing the integral, we clarify the point of the
integral range of modulation frequency from an experimental point of
view. In actual experiments a high-frequency cutoff is necessary.
In order to discuss the needed cutoff, let us recall the
setup of the Hubbard model.
Atoms in an optical lattice form a Bloch band structure, and the band
gaps are determined by depth of the optical lattice potential.
Since the single-band Hubbard model $H_0$ is introduced to demonstrate
the physics in the lowest Bloch band, the high energy cutoff necessary
to justify the effective model should be taken to be sufficiently small
compared with the band gap between the lowest and the next Bloch band.
Therefore, by integrating the DPR spectrum over the frequency region
below the band gap, the integration discussed below can be estimated in
experiments.

The integrated DPR~(\ref{eq:DPR-formula}) reads
\begin{equation}
 \int_0^{\infty}\!\! d\omega\
 \omega \tilde{\chi}^{\mathrm{R}}_{S}(\omega)
 = -\frac{1}{\hbar}
    \int_0^{\infty}\!\! dt\
      \frac{\langle[[H_0,S(t)],S(0)]\rangle_0}{t+i0^{+}}.
 \label{eq:integral}
\end{equation}
Here the quantity
$\langle[[H_0,S(t)],S(0)]\rangle_0$ in
Eq.~(\ref{eq:integral}) is real.
Thus the imaginary part in the right-hand side of
Eq.~(\ref{eq:integral}) is caused by the denominator of the integrand.
Using the decomposition,
$1/(t+i0^{+})=\mathcal{P}\frac{1}{t}-i\pi\delta(t)$ where $\mathcal{P}$
denotes the principal value, one can take the imaginary part, and obtain the
dimensionless total spectrum weight $\Gamma$ as
\begin{equation}
 \Gamma
 \equiv
 \int_0^{\infty}\!\!\frac{d\omega}{(J/\hbar)^2}\ P_{\mathrm{D}}(\omega)
 =-\frac{1}{\Omega}\frac{\pi\langle[[H_0,S],S]\rangle_0}{2J^{2}U}.
 \label{eq:gamma}
\end{equation}

This formula is one of the central results of this paper.
Remarkably, Eq.~(\ref{eq:gamma}) is deduced from
Eq.~(\ref{eq:DPR-formula}) without any approximation,
and in addition the inhomogeneity effect of the system is fully taken
into account in the formula. 
It relates a dynamical quantity to a calculation of equilibrium
correlation functions. 
Such correlations are easily amenable to a computation via a large
number of numerical techniques such as QMC simulations, allowing us to make
direct contact between the measured DPR spectrum and various quantities
of the Hubbard model. 
One can thus, by comparing the experimental and theoretical results,
expect to determine hard to get parameters, such as the temperature (or
entropy). 

Furthermore, as we will show below this general correlation, the generic
formula~(\ref{eq:gamma}) is reduced in the interesting regime of
temperatures $k_{B}T\ll U$ to a direct measure of the spin-spin
correlations. 
Equation~(\ref{eq:gamma}) contains two-particle correlation functions
such as
$\langle{c^{\dagger}_{i\uparrow}c^{\dagger}_{j\downarrow}c_{k\uparrow}c_{l\downarrow}}\rangle_0$.
Thus the practical calculation is not trivial in general, and the physical
interpretation is not also clear.
In what follows, we separately consider the two limited but interesting
regions: (i) high-temperature $k_{B}T\gg J$ and
(ii) strongly correlated regime $J\ll U$ at half-filling.
Note that the two-particle correlation functions can then be written as
\begin{align}
 \langle
   c^{\dagger}_{i\uparrow}c^{\dagger}_{j\downarrow}
   c_{k\uparrow}c_{l\downarrow}
 \rangle_0
 &=\delta_{i,k}\delta_{j,l}
    \langle
      c^{\dagger}_{i\uparrow}c^{\dagger}_{j\downarrow}
      c_{i\uparrow}c_{j\downarrow}
     \rangle_0
 \nonumber \\
 &\qquad
  +\delta_{i,l}\delta_{j,k}
    \langle
      c^{\dagger}_{i\uparrow}c^{\dagger}_{j\downarrow}
      c_{j\uparrow}c_{i\downarrow}
     \rangle_0.
 \label{eq:simplicification}
\end{align}

In the first case (i), hopping is incoherent regardless of the filling.
Thus we can simply apply Eq.~(\ref{eq:simplicification}) to the generic
formula~(\ref{eq:gamma}).
Consequently $\Gamma$ simplifies as
\begin{align}
 \Gamma
 &=-\frac{2\pi(\delta{F})^2}{\Omega}
    \left[
      \bar{N}_{\mathrm{D}}
      +\frac{1}{z}
       \sum_{\langle{i,j}\rangle}
        \left(
          \langle{\bm{S}_i\cdot\bm{S}_j}\rangle_0
          -\frac{\langle{n^{\mathrm{c}}_{i}n^{\mathrm{c}}_{j}}\rangle_0}{4}
        \right)
    \right]
 \nonumber \\
 &\quad
 -\frac{\pi(\delta{F})(\delta{J})}{2\Omega}
  \sum_{\langle{i,j}\rangle}
  \frac{v_i-v_j}{U}
  \left(
   \langle{n^{\mathrm{c}}_{i}}\rangle_0
   -\langle{n^{\mathrm{c}}_{j}}\rangle_0
  \right)
 \label{eq:imaginary-part-2}
\end{align}
where $n^{\mathrm{c}}_{j}=\sum_{\sigma}n_{j\sigma}$, and $z$ is a
coordination number.
The total doublon number in equilibrium state,
$\bar{N}_{\mathrm{D}}=\sum_{j}\langle{n_{j\uparrow}n_{j\downarrow}}\rangle_0$,
has also been introduced.
Note that the inhomogeneity of a system is also fully taken into account
in formula~(\ref{eq:imaginary-part-2}).
In the homogeneous case, since $v_i-v_j=0$, $\Gamma$
turns out to be given only by the first line.

In the second strongly interacting regime (ii), which allows us to reach
much lower temperatures, one has to restrict the study to the
half-filling homogeneous case to avoid regions in which coherent hopping
could still exist. 
In particular, in the presence of a trap, this will occur in the shell of
the compressible region. 
Hopefully such regions would give small contributions compared to the
bulk of the response.  
In addition, the response of these regions is mostly concentrated at low
energy rather than for energy or order $U$, and thus most of their
contribution can be filtered away in the frequency integration. 
If we restrict to the commensurate case, the
simplification~(\ref{eq:simplicification}) is still applicable 
for $k_{B}T \ll U$. Then the on-site potential term $H_{\mathrm{p}}$
essentially disappears~\footnote{It is easily found that
result~(\ref{eq:total-spectrum-weight}) in this discussion does not
change, even if we leave $H_{\mathrm{p}}$ in the form of a homogeneous
potential $v_j=v$.},
and the integrated spectrum $\Gamma$ turns out to be
Eq.~(\ref{eq:gamma}) in which $S$ is replaced by
$(\delta{F})H_{\mathrm{K}}$.
The resultant $\Gamma$ after applying
Eq.~(\ref{eq:simplicification}) is given by the first line of
Eq.~(\ref{eq:imaginary-part-2}).
However, the density fluctuation would then be strongly suppressed, and
one can approximately take $\bar{N}_{\mathrm{D}}\approx 0$
and $\langle{n^{\mathrm{c}}_{i}n^{\mathrm{c}}_{j}}\rangle_0\approx 1$,
whose temperature dependency would be negligibly small.
Therefore, the integrated DPR spectrum is given as
\begin{equation}
 \Gamma
 =2\pi z(\delta{F})^2
  \left[\frac{1}{4}
    -\frac{1}{z\Omega}
     \sum_{\langle{i,j}\rangle}
     \langle{\bm{S}_i\cdot\bm{S}_j}\rangle
  \right],
 \label{eq:total-spectrum-weight}
\end{equation}
where $z$ is a coordination number.
What is important is that only the nearest-neighbor spin-correlation
function $\langle{\bm{S}_{i}\cdot\bm{S}_{j}}\rangle$ is a dominant
contribution to the temperature dependence of $\Gamma$.
Namely, $\Gamma$ can be identical to the nearest-neighbor spin
correlation.

The nearest-neighbor spin correlation appearing in
Eq.~(\ref{eq:total-spectrum-weight}) is exactly the same as the energy
of the quantum spin Heisenberg model.
As is well known, in the strongly interacting region $k_{B}T,\, J\ll U$,
at half-filling, the Heisenberg model is deduced from the Hubbard model
as a consequence of the second-order perturbation theory in terms of
$J/U$. Thus, in such a region, Eq.~(\ref{eq:total-spectrum-weight})
allows us to probe the system energy.

In order to see the temperature dependence of $\Gamma$, we
numerically estimate the temperature dependence of the integrated DPR
spectrum $\Gamma$. 
However, it is a highly nontrivial problem because it is necessary to
compute the dynamical quantity~(\ref{eq:DPR-formula}) directly in a wide
regime of temperature.
Thus, in this paper the parameter region is restricted to the regime for
which the approximation~(\ref{eq:total-spectrum-weight}) is valid,
i.e., $k_{B}T,\, J\ll U$.
In addition, we calculate the static quantity (nearest-neighbor
spin correlation)
$\langle{\bm{S}_{i}\cdot\bm{S}_{j}}\rangle$ instead of
integrating the DPR spectrum over the modulation frequency.
In the low-temperature Mott insulating case, we can approximately
choose the AFM Heisenberg model as an effective model,
which allows one to compute the nearest-neighbor spin correlation
through the energy of the Heisenberg model.

\begin{figure}[t]
 \includegraphics[scale=.7]{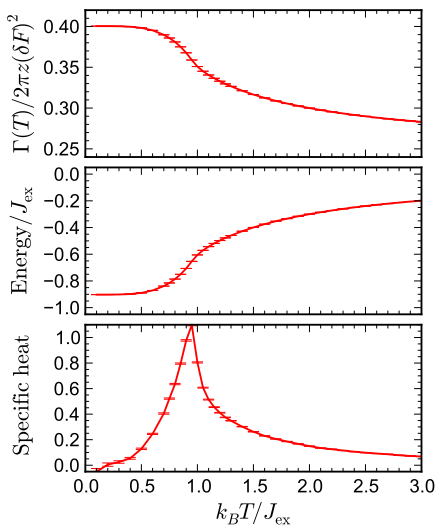}
 \caption{(Color online) The temperature dependence of (top) the
 integrated DPR spectrum  per site $\Gamma$ scaled by
 $2\pi z(\delta{F})^2$,
 (middle) the energy per site scaled by a superexchange spin
 coupling $J_{\mathrm{ex}}$, which is identical to the nearest-neighbor
 spin correlation per site multiplied by the coordination number $z=6$
 in a cubic lattice, and
 (bottom) the specific heat.
 The QMC calculation is implemented for the quantities in the Heisenberg
 model for $14^3$ system size.
 While error bars are also shown, they are negligibly small.
 The N\'{e}el transition occurs near $k_{B}T\approx J_{\mathrm{ex}}$,
 and then the integrated spectrum and the energy per site
 exhibit an inflection-point-like behavior.
 }
 \label{fig:Gamma_of_temperature}
\end{figure}

We implement the QMC calculation for the AFM Heisenberg model of a cubic
lattice where the system size is $\Omega=14^3$~\footnote{The
calculations for $\Omega=12^3$ and $10^3$ have also been implemented,
but no crucial difference has been found on the curves, showing that
finite size corrections are already very small.}.
The result is shown in Fig.~\ref{fig:Gamma_of_temperature}, where the
calculated energy is translated into $\Gamma$ by use of the
relation~(\ref{eq:total-spectrum-weight}).
As a reference of the critical temperature of the N\'{e}el
transition~\cite{Wessel/PRB81.2010,Sandvik/PRL80.1998},
the specific heat as a function of temperature is also shown.
The derivative of $\Gamma$ as temperature rises is found to be
maximum near the critical point, which seems an inflection point.
On the other hand, the nearest-neighbor spin correlation and the energy
for temperatures higher than the critical point exhibit long tail
decay. 

To summarize, we have explicitly discussed the DPR spectra in
the presence of an inhomogeneous potential, and the generalized form of
the DPR spectrum as a function of lattice modulation frequency has been
obtained. 
In addition we have implemented the integration of the DPR spectrum over
modulation frequency.
The problem of dynamics can be reduced to the calculation of
static quantity as shown in Eq.~(\ref{eq:gamma}). 
Such a formula can be the basis of a very accurate numerical comparison
with the experiment.
Furthermore, focusing on the two different regimes, (i) $k_{B}T\gg J$
and (ii) $J\ll U$,
a generic form of the integrated DPR spectrum has been deduced.
In particular, for the incompressible Mott regime, it is found to be
associated with the nearest-neighbor spin correlation as in
Eq.~(\ref{eq:total-spectrum-weight}), and can thus be used as a probe of
the magnetic properties. 
In the larger $J$ regime, higher order hopping processes would be more
relevant, and the integrated DPR spectrum is expected to probe
longer-range spin correlations.
One can thus use the shaking of the lattice and DPR to directly probe
for magnetic order.

\acknowledgments
The calculation for temperature dependence of the energy has been
implemented by using the looper QMC code of the ALPS libraries~\cite{Todo.Kato/PRL87.2001/ALPS.looper,Evertz.Lana.Marcu/PRL70.1993/ALPS.looper,Beard.Wiese/PRL77.1996/ALPS.looper,Evertz/AdvPhys52.2003/ALPS.looper.review,Albuquerque.etal/JMagMagMat310.2007/ALPS,Bauer.etal/JStatMechP05001.2011/ALPS}.
This work was supported by the Swiss National Foundation under MaNEP and
division II.

\bibliographystyle{apsrev4-1}
%% \bibliography{reference}
%

\end{document}